# Using XDAQ in Application Scenarios of the CMS Experiment

V. Brigljevic, G. Bruno, E. Cano, A. Csilling, S. Cittolin, D. Gigi, F. Glege, M. Gulmini[1], J. Gutleber*, C. Jacobs, M. Kozlowski, H. Larsen, I. Magrans, F. Meijers, E. Meschi, L. Mirabito, S. Murray, A. Oh, L. Orsini, L. Pollet, A. Racz, D. Samyn, P. Scharff-Hansen, P. Sphicas[2], C. Schwick

*CERN, European Organization for Nuclear Research, Geneva, Switzerland*
[1]*Also at Laboratori Nazionali di Legnaro, INFN, Legnaro, Italy*
[2]*Also at University of Athens, Greece*

F. Drouhin
*Universite de Haute-Alsace, Mulhouse-France - Institut de Recherche Subatomique de Strasbourg, France*

L. Berti, G. Maron, N. Toniolo, L. Zangrando
*INFN, Laboratori Nazionali di Legnaro, Legnaro, Italy*

S. Ventura
*INFN - Sezione di Padova, Padova, Italy*

S. Erhan
*University of California, Los Angeles, California, USA*

V. O' Dell, I. Suzuki
*Fermi National Accelerator Laboratory, Batavia, Illinois, USA*

*presented by Johannes Gutleber (Johannes.Gutleber@cern.ch)

XDAQ is a generic data acquisition software environment that emerged from a rich set of of use-cases encountered in the CMS experiment. They cover not the deployment for multiple sub-detectors and the operation of different processing and networking equipment as well as a distributed collaboration of users with different needs. The use of the software in various application scenarios demonstrated the viability of the approach. We discuss two applications, the tracker local DAQ system for front-end commissioning and the muon chamber validation system. The description is completed by a brief overview of XDAQ.

## 1. INTRODUCTION

Software for data acquisition (DAQ) systems does not only comprise the application task, but also requires functions to integrate diverse hardware devices for configuration and data interchange. With traditional tools this can become a cumbersome and time-consuming task. Application software that relies on direct use of the device driver interface affects program configurability. Some devices are accessed through system calls, others rely on memory mapping or have special requirements like the existence of memory that can be used for DMA operations. Support for configuration is dependent on the operating system platform. Concurrent use of multiple different networking technologies transparently at application level is not easily achievable with this approach. Message passing libraries, such as MPI [1] provide communication abstraction but do merely support the integration of diverse technologies. Moreover, different tasks in high-energy physics data acquisition show striking similarities although the underlying networking and processing devices can be entirely different. Recognition of these limitations triggered an abstraction from a pure application-oriented view of data acquisition. Related projects showed that this direction is promising. CODA [2] for example, presents an integrated data acquisition environment. Highly portable and feature rich, it is, however, limited in terms of integration with other systems because of proprietary protocols and formats.

We approached the problem by filtering out generic requirements that are common to various tasks in high-energy physics [3]. We then established a software product line [4], specifically designed for distributed data acquisition systems based on the integration of various networking devices and commodity computing systems. This suite, called XDAQ, includes the generic requirements documents, design templates, a software process environment, a distributed processing environment and various generic software components that can be tailored to a variety of application scenarios (see figure 1). Applying the product line approach to data acquisition aims, however, at shifting the focus from application programming tasks to integration tasks, thus speeding up application development and obtaining good performance by using well-established and tested design patterns. Before we present the use of XDAQ in some of the application scenarios of the CMS experiment, we outline basic functionalities of the software environment.

## 2. XDAQ

XDAQ is a software product line that has been designed [5] to match the diverse requirements of data acquisition application scenarios of the CMS experiment. These include the central DAQ, sub-detector local DAQ systems for commissioning, debugging, configuration, monitoring and calibration purposes, test-beam and detector production installations as well as design verification and demonstration purposes.

**MOGT008**

The product line comprises sets of documentation and software packages that are generic enough to be used for several application scenarios, but as specific as possible to cover aspects that are critical to stable and efficient DAQ operation.

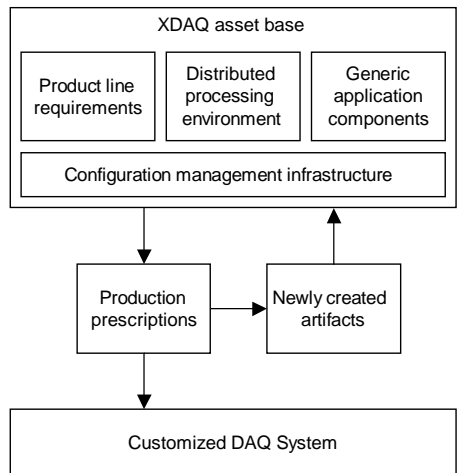

Figure 1: Overview of the XDAQ software product line. The arrows indicate the work flow of the software process that leads to the production of DAQ systems and the extension of the product line asset base.

### 2.1. Documentation

The documentation includes besides a description of the project environment, design documents, software production guidelines, user- and test manuals a generic requirements specification. It captures requirements that are common to a large set of data acquisition systems, which aim to use commodity computing and networking equipment. At functional level the requirements cover communication, configuration, monitoring and control tasks. In addition, various non-functional requirements like performance flexibility, maintainability and portability are included. This description must be completed by the requirements that are specific to a particular application scenario and will lead to tailoring of the generic application components that are provided with the suite. Some upcoming requirements may eventually be found in multiple environments and can eventually end up in the product-line requirements specification.

A second important documentation set is the collection of companion manuals. It serves as a production guide, telling a user how to build a specific data acquisition application by using the generic processing environment and the included application components. Some manuals focus on tailoring aspects, i.e. how to augment the existing components with the additional functionalities required by the application. Others give insight in using existing functions for the necessary programming tasks.

### 2.2. Distributed Processing Environment

XDAQ includes a distributed processing environment called "the executive" that provides applications with the necessary functions for communication, configuration control and monitoring [5, 6]. Written entirely in C++ with an emphasis on platform independence, it implements well-established techniques to provide applications with efficient, asynchronous communication. They include the use of memory pools for fast and predictable buffer allocation [7], support for zero-copy operation [8, 9] and an efficient dispatching mechanism for an event-driven processing scheme [10]. A copy of the executive process runs on every processing node in the data acquisition network. Applications are modeled according to a software component model [11] and follow a pre-scribed interface. They are compiled and the object code is loaded dynamically, at run-time into a running executive. Multiple application components, even of the same application class may coexist in a single executive process. All configuration, control and monitoring can be performed through the SOAP/http [12] protocol, widely used in Web enabled applications [13]. A rich set of data structures, including lists, vectors are exportable and can be inspected by clients through the executive SOAP services. Histograms are mapped to data structures, too. They can also be retrieved via SOAP.

### 2.3. Generic Event Builder

In addition to documentation and the executive, XDAQ includes a collection of generic applications. They are ready for use in various application scenarios with tailoring points that allow the adaptation to specific environments. One of them is an event builder [14] that consists of three collaborating components, a readout unit (RU), a builder unit (BU) and an event manager (EVM). The logical components and interconnects of the event builder are shown schematically in figure 2. A summary of the acronyms is given in table 1. Data that are recorded by custom readout devices are forwarded to the readout unit application. How this is accomplished is described in an associated document that is provided with a template software module. A RU buffers data from subsequent single physics events until it receives a control message to forward a specific event fragment to a builder unit. A builder unit collects the event fragments belonging to a single collision event from all RUs and combines them to a complete event. The BU exposes an interface to event data processors, called the filter units (FU). This interface can be used to make event data persistent or to apply event-filtering algorithms. The EVM interfaces to the trigger readout electronics and so controls the event building process by mediating control messages between RUs and BUs. The interface between EVM and trigger readout is documented and template code for adaptation to various custom devices is provided.

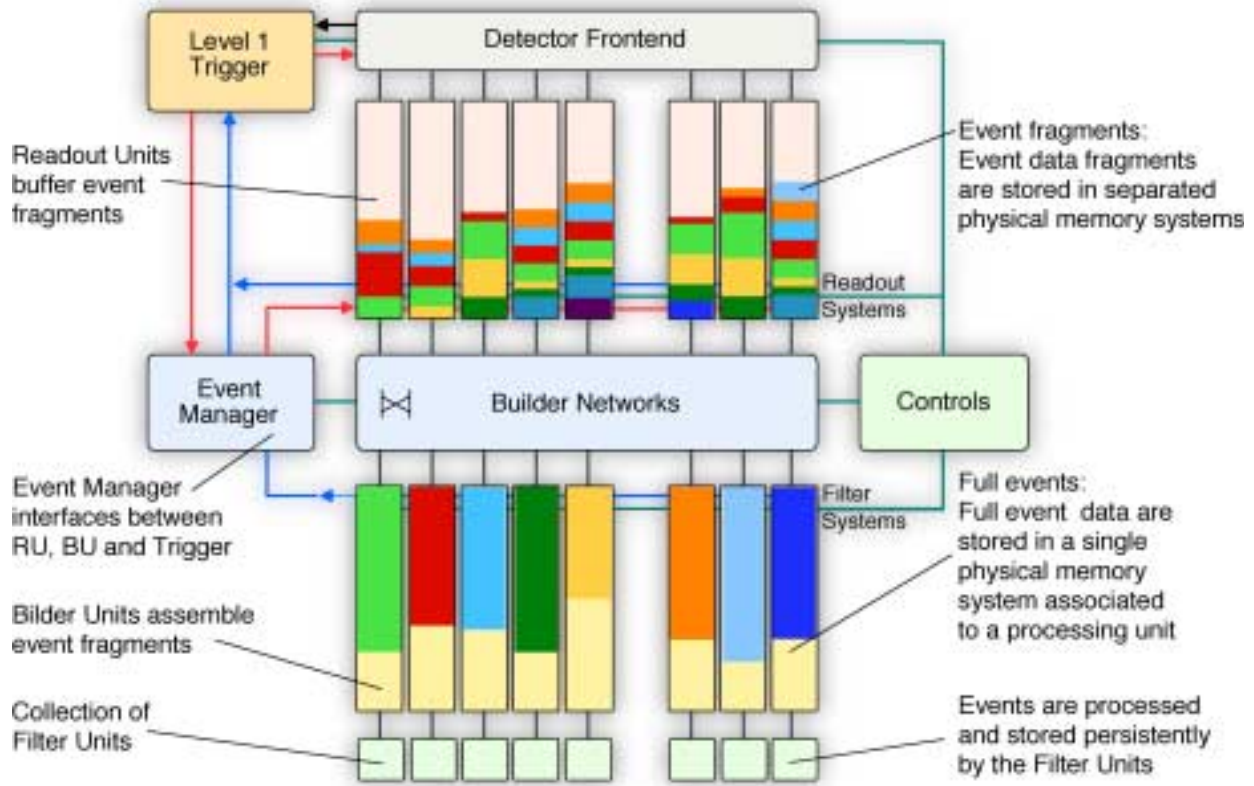

Figure 2: Outline of an event building data acquisition system.

## 3. APPLICATION SCENARIOS

We outline two of the various DAQ application scenarios found in the CMS experiment. The first one is the silicon microstrip detector commissioning of the tracker sub-detector. The second use case is the test and validation of the muon chambers. The system overviews include descriptions of the software and hardware subsystems, including the interconnection technologies.

Table 1: List of used acronyms.

| Acronym | Description |
|---|---|
| BCN | *Builder Control Network,* a logical network used to communicate requests for event data from BU to EVM. |
| BDN | *Builder Data Network,* logical network used to send event data from RU to BU |
| BU | *Builder Unit,* hardware/software subsystem to collect event data from RUs |
| DCS | *Detector Control System*, software system to control/monitor high and low voltages as well as various environmental measurement values |
| DSN | *DAQ Service Network*, logical network to carry run control and DCS messages |
| EVM | *Event Manager*, hardware/software subsystem to mediate trigger information for readout and event building purposes as well as to control the event building process |
| FEC | *Front-End Controller*, mediates control information between custom front-end devices, the DCS or run-control/monitor system |
| FED | *Front-End Driver*, custom devices to read data from the detector elements |
| FFN | *Filter Farm Network*, logical network transports event data from BUs to FUs |
| FRL | *Front-End Readout Link*, interconnect to read event data from FED into a RU. |
| FU | *Filter Unit*, hardware/software subsystem to process and store event data |
| LTC | *Local Trigger Controller*, subsystem for reading trigger information and for controlling the trigger. |
| RCMS | *Run Control/Monitor System*, software subsystem to configure, control and monitor the DAQ system |
| RCN | *Readout Control Network*, logical network to transport control information for tagging events in a RU and for forwarding the data to BUs |
| RU | *Readout Unit*, hardware/software subsystem for transient storage of event data read from one or more FEDs |

## 3.1. TRACKER COMMISSIONING

The CMS tracker [15] comprises silicon microstrip detectors [16] that are read out via an analog optical link connected to the analog to digital converters of the FED. The control commands for the front-end chips, together with the clock and level 1 trigger, are propagated to the detectors through a proprietary token-ring network. This task is carried out by a front-end controller (FEC) interacting with several communication and control units. Commissioning of detectors is a dedicated data acquisition task including the communication channel described above. For example, timing calibration is required, because the trigger is propagated sequentially along the ring to the detectors and the digitization time depends on the fiber lengths to the FEDs. Additional calibration tasks include pulse shape adjustment, optical link gain setting and front-end gain calibration. Test setups include slow control facilities, such as thermistors and $I^2C$ driven humidity probes, and HV/LV control, which are driven by XDAQ applications. For this purpose, a local data acquisition system that supports the calibration loop process in addition to configuration and control operations must be available. A high-level diagram of the system in operation is shown in Figure 3. Subsystem implementations are listed in Table 2. XDAQ together with generic event builder components has been successfully used to implement the system described above. Additional specialized software components were developed to interface to detector specific electronics and various persistent data storage technologies. On-line visualization facilities were implemented with Java Analysis Studio [17] and interfaced to the system through the SOAP messaging system. The implementation of the system took 4 man months. Flexibility and scalability of XDAQ was demonstrated by its use in different configurations. It was possible to transfer an existing small setup from PSI (Zurich, Switzerland) to tracker subsystem (rod and petal) tests without modifications. This system was easily transferred to a testbeam environment at CERN that comprised different computer resources. This setup comprised in total 8 computers: one for the EVM, three for the FUs, one for the BU, one for the RU (hosting three RU applications), one for the FEC and one for run control/detector control. Commissioning of the new system took 2 hours as opposed to 30 hours with previous systems. At a rate of 2000 events per spill (500 Hz with an average event size of 20 Kbytes), a maximum data throughput of 100 Mbit/s was achieved, which corresponds to the available Fast Ethernet capabilities between the RU and BU subsystems. Recently, an upgrade to Gigabit Ethernet has been performed, yielding a throughput of 71 Mbytes/sec for the data acquired during one spill. Operation periods spanned five days of continuous and uninterrupted data taking resulting in around 600 Gbytes of data produced for analysis. Novice users were able to operate and re-configure the distributed system after a training of approximately 1 hour by themselves. In particular the application of configuration changes to re-distribute processing tasks to multiple computers was achieved without further assistance.

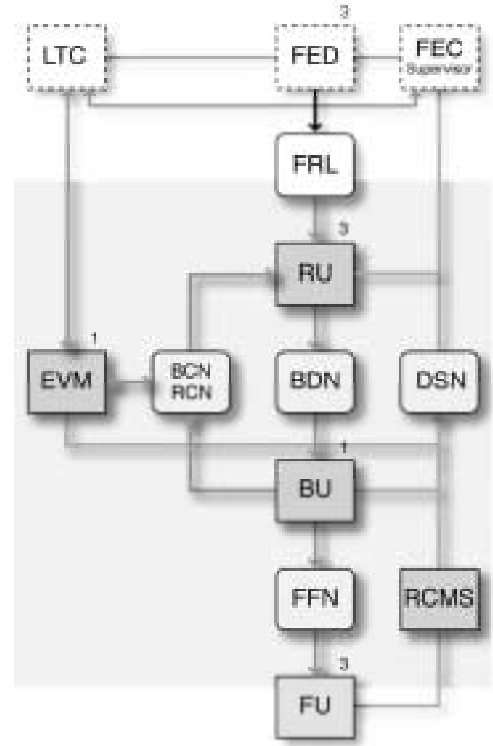

Figure 3: Overview of the tracker commissioning system

Table 2: Subsystem/interconnect implementation

| Subsystem | Implementation |
|---|---|
| BCN | Fast Ethernet ($I_2O$ binary messages) |
| BDN | Fast Ethernet ($I_2O$ binary messages) |
| BU | Intel based PC (Linux) |
| DCS | Custom (XDAQ based applications) |
| DSN | Fast Ethernet (SOAP/http) |
| EVM | Intel based PC (Linux) |
| FEC | Intel based PC (Linux) |
| FED | Custom PCI cards |
| FFN | Fast Ethernet ($I_2O$ binary messages) |
| FRL | PCI bus |
| FU | Intel based PC (Linux) |
| LTC | Custom PCI card |
| RCMS | Java based (xdaqWin, JAXM) |
| RCN | Fast Ethernet ($I_2O$ binary messages) |
| RU | Intel based PC (Linux) |

## 3.2. MUON Chamber Validation

The 250 chambers that make up the muon barrel spectrometer [18] need to be tested with cosmic rays once before they are shipped to CERN and once when they arrive. Tests with muon beams at high data rates together with coupling to other subdetectors are also necessary to validate the detector behavior under realistic conditions. The observables include occupancy of detector channels for determining the particle hits, the drift time and the response to dedicated test pulse events. A data acquisition system had to be put in place to perform these tasks. XDAQ was used as the generic platform to implement this system. As outlined in Figure 4, the software required customization of three parts, the local trigger controller (LTC), readout of custom electronics through the VME bus and interfacing to a data storage/analysis backend. For the acquisition task, the XDAQ generic event builder components were used. Run control was implemented by a prototype system that served as a study environment for the CMS experiment run control system [19]. As opposed to the tracker commissioning slow control was performed by a separate, Windows based system. A similar setup was used during a test period for the Gamma Irradiation Facility at CERN. In this case, a muon chamber was coupled to a resistive plate chamber (RPC) detector and a beam ionization chamber to verify the simultaneous response of the RPC and muon chambers. The update of the existing system went smoothly. As a result of this project, several new requirements were identified, mainly covering configuration, control and monitoring. Hardware platform heterogeneity (different bus systems, byte ordering and data alignment rules), as well as the presence of two different operating system platforms (Linux and VxWorks), posed a challenge to the interoperability among the system components. The supported platforms included PowerPC based VME processors running the VxWorks real-time operating system, and Intel based personal computers with both Linux and VxWorks.

Table 2: Subsystem implementation

| Subsystem | Implementation |
|---|---|
| BCN | Fast Ethernet ($I_2O$ binary messages) |
| BDN | Fast Ethernet ($I_2O$ binary messages) |
| BU | Intel based PC (Linux) |
| DSN | Fast Ethernet (SOAP/http) |
| EVM | Intel based PC (VxWorks) |
| FED | Custom VME cards |
| FRL | VME bus |
| LTC | Custom PCI card |
| RCMS | Java application (CMS prototype) |
| RCN | Fast Ethernet ($I_2O$ binary messages) |
| RU | Intel based PC (Linux and VxWorks) |

The provided abstraction fitted the need to switch between processor and operating system types without additional work. The required event rate of 10 kHz and the peak output of 4 Mbytes/sec were well absorbed by the hardware (PCs, VME CPUs and Fast Ethernet network) and software installation in place. The high variance of the event sizes stressed the buffer management system of XDAQ. Stable operation under these conditions confirmed the robustness of the design.

**MOGT008**

Uptime measured over two weeks was 60%, including penalties from changing hardware configurations. From the initial intent to create the system to its completion, including the learning phase, six man months were invested. A system developed from scratch in the same time scale would not have provided the seamless integration with later appearing components (e.g. silicon beam telescope) and the ability to efficiently carry out modifications and configuration changes. Through this experience, confidence has been gained that the proposed design of the software infrastructure can fulfill the diverse functional DAQ requirements of the experiment under design.

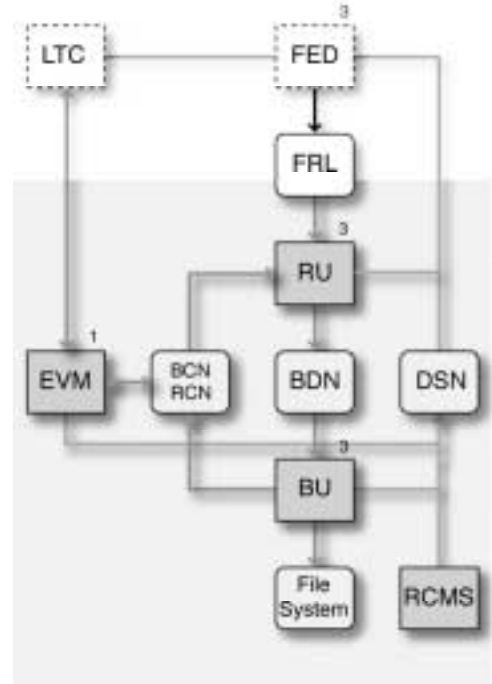

Figure 4: Overview of the Muon validation system.

## 4. SUMMARY

In this paper we outlined two application scenarios of XDAQ in the CMS experiment. The creation of DAQ systems for the tracker commissioning and the muon chamber validation tasks were vital to ensure that the online software infrastructure under design and implementation eventually meets the experiments requirements. XDAQ supports DAQ systems that aim to use commodity computing and networking equipment. It is also evaluated for use in the experiment's central DAQ system that favors this design choice. Performance and scalability studies for the main DAQ system are currently done [20]. The preliminary results from these studies show that the performance of the software matches well the requirements [21, section 2].